\documentclass[graybox,natbib,nosecnum]{svmult}
\bibpunct{(}{)}{;}{a}{}{,} % suppress commas between author-names and year

\pdfoutput=1   %forces use of pdflatex. Disable if you prefer to use .eps or .ps figures.
\usepackage{mathptmx}       % selects Times Roman as basic font
\usepackage{helvet}         % selects Helvetica as sans-serif font
\usepackage{courier}        % selects Courier as typewriter font
\usepackage{type1cm}        % activate if the above 3 fonts are
\usepackage{makeidx}         % allows index generation
\usepackage{graphicx}        % standard LaTeX graphics tool
\usepackage{multicol}        % used for the two-column index
\usepackage[bottom]{footmisc}% places footnotes at page bottom
\usepackage[normalem]{ulem}	% for strike-through of text with \sout{}  
\usepackage{hyperref}  %for hyperlinks

\usepackage{soul}   % for high-lighting of text
\makeindex             % used for the subject index
\begin{document}

\title*{Debris Disks: Probing Planet Formation}
\author{Mark C. Wyatt}
\institute{Mark C. Wyatt \at Institute of Astronomy, University of Cambridge, Madingley Road, Cambridge CB3 0HA, United Kingdom, \email{wyatt@ast.cam.ac.uk}}
\maketitle

\abstract{Debris disks are the dust disks found around $\sim 20$\% of nearby main sequence
stars in far-IR surveys.
They can be considered as descendants of protoplanetary disks or components of
planetary systems, providing valuable information on circumstellar disk evolution and the
outcome of planet formation.
The debris disk population can be explained by the steady collisional erosion
of planetesimal belts;
population models constrain where (10-100\,au) and in what quantity
($>1M_\oplus$) planetesimals ($>10$\,km in size) typically form in
protoplanetary disks.
Gas is now seen long into the debris disk phase.
Some of this is secondary implying planetesimals have a Solar System comet-like
composition, but some systems may retain primordial gas.
Ongoing planet formation processes are invoked for some debris disks,
such as the continued growth of dwarf planets in an unstirred disk,
or the growth of terrestrial planets through giant impacts.
Planets imprint structure on debris disks in many ways;
images of gaps, clumps, warps, eccentricities and other disk asymmetries,
are readily explained by planets at $\gg 5$\,au.
Hot dust in the region planets are commonly found ($<5$\,au) is seen for
a growing number of stars.
This dust usually originates in an outer belt (e.g., from exocomets),
although an asteroid belt or recent collision is sometimes inferred.
}
%The abstract should be 10-15 lines (250 words).
%The paper should be up to, and typically, 15 pages long (incl references).
%Highlight \hbindex{relevant terms} for the Handbook's index.

%%%%%%%%%%%%%%%%%%%%%%%%%%%%%%%%%%%%%%
\section{Introduction}
Many 100s of debris disks are now known, most of which were discovered in far-IR surveys of nearby stars
implying the presence of cold dust.
The ubiquity of debris disks is apparent from the fact that such circumstellar dust is
present at a detectable level around $\sim 20$\% of nearby main sequence stars.
The dust is thought to have a much shorter lifetime than the stellar age, implying that this is not
primordial (i.e., not left over from the protoplanetary disk phase). 
Rather the dust is inferred to be secondary, continually replenished from the break-up of larger planetesimals.

The proximity to the Sun of many debris disks means the structure of this cold dust can be studied at high resolution.
Such imaging has shown that the dust is often confined to a ring-like geometry at 10s of au.
Thus it is tempting to interpret debris disks as extrasolar analogues to the Kuiper belt.
A smaller fraction of stars show evidence in mid-IR observations for warm dust at a few au,
which may be analogous to the Solar System's zodiacal cloud.  
Implicit in these analogies is the existence of a putative planetary system in the regions
absent of dust.
However, for most stars there is no strong evidence for the presence of any planets,
and it remains possible that the parent planetesimals reside in otherwise empty systems.
Nevertheless, given the ubiquity of extrasolar planets, and the growing number of systems with both
debris and planets, it is plausible that {\bf a debris disk is just one component of an underlying
planetary system}.

This view of planetary systems as emerging fully formed from the protoplanetary disk phase
shows that {\bf observations of the debris disk component provide valuable information on 
the outcome of planet and planetesimal formation processes}.
The presence of dust indicates where planetesimals either formed or ended up,
and the detailed structure of the disks give clues to where there may be planets.
However, again taking the example of the Solar System, it is also clear that planetary systems do not
necessarily remain static after protoplanetary disk dispersal.
The Earth's moon is thought to have formed in a giant impact at 50-100\,Myr, and a dynamical instability involving
the giant planets is thought to have lead to a restructuring of the planetary system architecture
and depletion of the Kuiper and Asteroid belts possibly as late as 700\,Myr after the formation
of the Solar System.
Both of these events would have been accompanied by a significant change in the observable properties
of the Solar System's debris disk.
Thus {\bf debris disks can also provide evidence of ongoing planet formation processes}.

While the dust seen in debris disks may not be primordial, it is nevertheless still helpful to
consider that {\bf debris disks are the descendants of protoplanetary disks, informing on the processes
of protoplanetary disk dispersal}.
This is true for the dust because the solid material seen in the earlier phases must go somewhere,
and there may be planetesimals present already within the protoplanetary disk.
This view is also particularly relevant when considering the gaseous component of debris
disks, evidence for which is accumulating for many stars, some of which are inferred to be have
a component of primordial gas.

%%%%%%%%%%%%%%%%%%%%%%%%%%%%%%%%%%%%%%
\section{Debris disks as descendants of protoplanetary disks}
\label{s:ppd}
\begin{figure}
%\vspace{-0.8in}
%\hspace{-0.9in}
\includegraphics[scale=0.60]{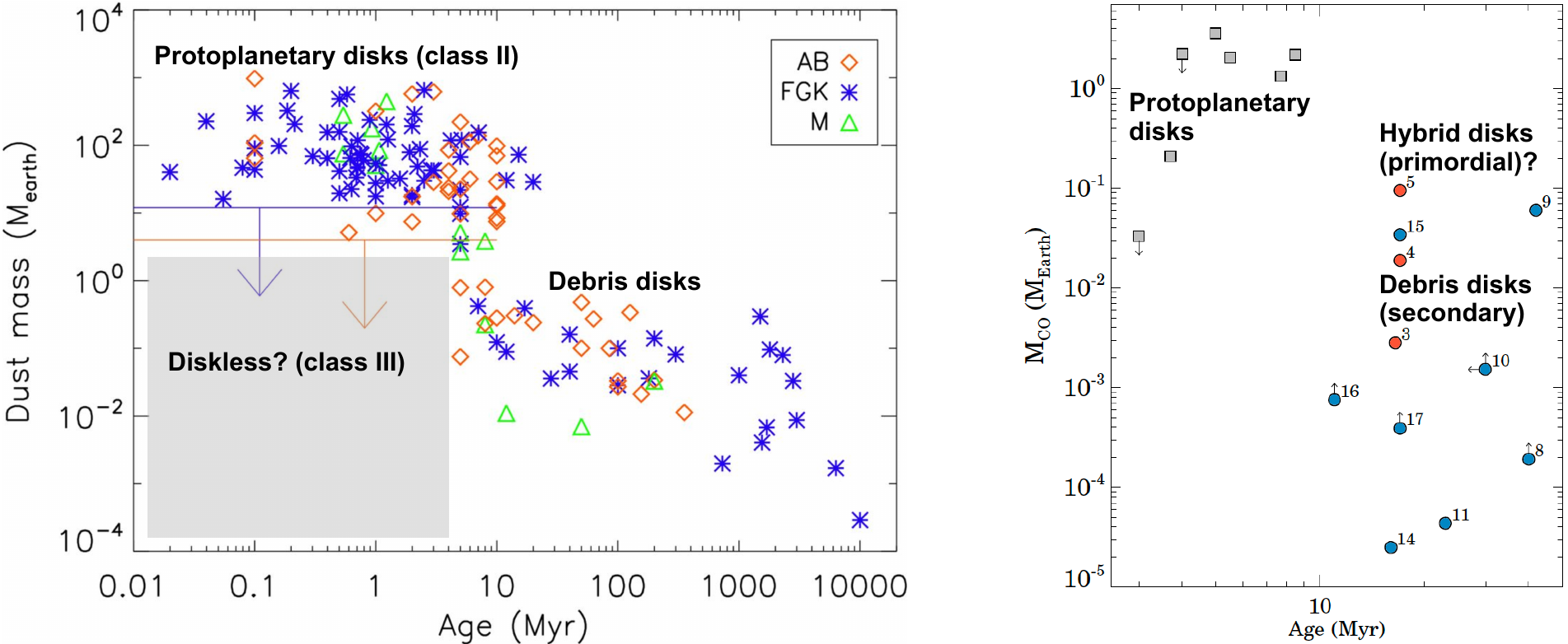}
%\vspace{-1.5in}
\caption{Evolution of circumstellar disk mass, as determined from observations of sub-mm dust \citep[left,][]{Panic2013}
and CO gas \citep[right,][]{Moor2017}.
Debris disks and protoplanetary disks are readily distinguished by the difference in disk mass and age
on both plots.
Horizontal lines on the left plot show the approximate pre-ALMA detection threshold in nearby star forming regions,
which was insufficient to detect debris disk levels of dust around class III stars.
Debris disk gas is likely to be secondary, although large CO gas masses around some young stars could
indicate a primordial component, earning such disks the name {\it hybrid} disks.
}
\label{fig:1}       % Give a unique label
\end{figure}

In contrast to protoplanetary disks, which are optically thick at optical and infrared wavelengths
and have large masses in mm-sized dust ($>1M_\oplus$) and
even more mass in gas, debris disks are optically thin with low mm-sized dust masses ($<0.1M_\oplus$)
and very little if any evidence for gas (see Fig.~\ref{fig:1}).
However, the boundary between the two types of disk is not well defined, and
there are several examples of disks with different classifications depending on the criteria used
\citep[e.g.,][]{Schneider2013,Kennedy2014}.
Part of the problem arises because it is not possible to detect analogues to the brightest debris disks
(that are within a few 10s of pc) in nearby star forming regions that are beyond 100\,pc \citep[e.g.,][]{Hardy2015};
e.g., stars in such regions that do not show evidence of 24\,$\mu$m emission above photospheric
levels are designated as class III stars (as opposed to class I or II stars for which excess 24\,$\mu$m
emission is present), and are normally considered diskless, even though these could harbor debris disks for
which the dust emission simply lies below the detection threshold \citep[e.g.,][]{Cieza2013}.
Thus it has been challenging to piece together the mechanism of protoplanetary disk dispersal and the
subsequent (or concurrent, or indeed prior) birth of a debris disk.

\citet{Wyatt2015} outlined 5 steps in the transition from protoplanetary to debris disk.
The first of these is the carving of an inner hole resulting in a morphology known as a transition disk,
which may be evident from a disk image by an absence of dust emission in the regions close to the star,
or from an absence of hot dust in the emission spectrum.
However, it should be noted that it remains debated whether this transition disk morphology is a necessary
step in the transition or just a different class of protoplanetary disk.
The remaining 4 steps involve the removal of cold mm-sized dust from the outer 10s of au, the emptying of
dust from the inner regions, the dispersal of the gas disk, and the shepherding of planetesimals into ring-like
structures.
It was acknowledged that the order of these steps is unknown and may vary between systems.
It was also concluded that a defining feature of protoplanetary disks is the presence of gas in sufficient
quantities to affect the dynamical evolution of the dust, and it is the gaseous component that will 
be the focus of this section.

%%%%%%%%%%%%%%%%%%%%%%%%%%%%%%%%%%%%%%%%%%%%%%%%%%%%%%%%%%%%%%
\subsection{Debris disk gas}
The last few years have seen significant progress in our understanding of the evolution of circumstellar
gas between the phases.
Most importantly it is now clear that a gaseous component is present long into
the debris disk phase for many systems.
For example, whereas in the pre-ALMA era absorption by circumstellar gas along the line of sight was seen
to two stars with edge-on debris disks, thanks to ALMA we now know that detectable levels of CO gas emission are present
around roughly half (11/17) of 10-40\,Myr-old A stars within 150\,pc that have debris disks with fractional
luminosities in the range 0.05-1\% \citep{Lieman-Sifry2016,Greaves2016,Moor2017};
here fractional luminosity is defined as the ratio of the luminosity of the dust emission to that of the star.
Gas mass estimates contain many uncertainties, but the strength
of the CO lines is diminished for debris disks compared with protoplanetary disks
\citep[e.g.,][]{Pericaud2017}, although for some debris disk systems the inferred
disk mass may be comparable (see Fig.~\ref{fig:1}).
For now it is young A stars that dominate the samples of debris disk gas detections \citep{Lieman-Sifry2016},
but gas is also seen around both later F-type stars \citep{Marino2016,Marino2017}
and older stars up to 1\,Gyr \citep{Matra2017b,Marino2017}.

%%%%%%%%%%%%%%%%%%%%%%%%%%%%%%%%%%%%%%%%%%%%%%%%%%%%%%%%%%%%%%
\subsubsection{Secondary origin of debris disk gas}
While the ubiquity of circumstellar gas is becoming clear, its origin remains unknown for most systems.
The gas could be either primordial (i.e., a remnant of the protoplanetary disk) or secondary
(i.e., the volatiles are locked in icy planetesimals until released in collisions).
Differentiating between these possibilities is difficult because (similar to protoplanetary disks)
it is still hard to detect anything but CO.
Nevertheless, for some systems an unambiguous origin for the gas can be ascertained.
For example, for $\beta$ Pictoris the striking similarity of the morphology of the CO gas disk and that of
the dust disk, both of which are distributed over 60-130\,au and include a prominent clump,
argues for a secondary origin \citep{Dent2014}.
This is supported by the CO line ratios that rule out a primordial abundance of hydrogen which if present
would lead to LTE line ratios that can be ruled out \citep{Matra2017}.

That we can detect CO at all in $\beta$ Pictoris is inferred to arise from a high CO production rate of $0.1M_\oplus$/Myr,
since photodissociation by interstellar radiation occurs on 120\,yr timescales.
It is hard to get around this short CO lifetime, even with CO self-shielding, and is the likely
explanation for the small numbers of CO detections around other debris disks until now.
The dust production rate is also high in this disk, and assuming that gas and dust are lost in proportion
to their composition, an up to 6\% by mass ratio of CO is inferred consistent with Solar System comets
\citep{Matra2017}.
This picture of steady state replenishment of the gas is reinforced by the detection of the atomic
C disk by far-IR spectroscopy of the CII line at levels compatible with this scenario if the
atomic gas spreads viscously \citep{Kral2016}.
This is also compatible with observations of OI \citep{Brandeker2016}, particularly if the planetesimal
composition includes H$_2$O in similar quantities to Solar System comets \citep{Kral2016}.
The situation is similar in several other debris disks with CO gas detections;
e.g., 23\,Myr HD181327 and 400\,Myr Fomalhaut both have CO coincident with their much narrower dust rings
at levels compatible with steady state production from comets with of order 10\% volatile fraction \citep{Marino2016,Matra2017b}.
In fact, applying this logic to all known dusty debris disks shows that it is plausible that all of these
have comet-like compositions and hence also have CO, CII, OI and CI present at some level, with only the brightest
systems detected thus far \citep{Kral2017}.

%%%%%%%%%%%%%%%%%%%%%%%%%%%%%%%%%%%%%%%%%%%%%%%%%%%%%%%%%%%%%%
\subsubsection{Primordial debris disk gas: hybrid disks?}
The success of a secondary explanation for the $\beta$ Pictoris gas disk is in contrast to its failure to explain
the gas seen toward HD21997 \citep{Kospal2013}.
The CO gas orbiting this 30\,Myr A3V star is found at radial locations (from $<26$\,au to 138\,au) where there are no
mm-sized dust grains (which are found from 55 to $>150$\,au and thought to trace the location of planetesimals). 
The short gas lifetime requires the two to be co-located in the secondary scenario, leading to the
suggestion that the gas is primordial, or that this is a {\it hybrid} disk where the gas has both origins.
However, a short CO lifetime is equally problematic for the primordial scenario, unless the protoplanetary disk
has dispersed implausibly recently.
This suggests that the CO is shielded somehow to prevent its photodissociation.
While the $^{12}$CO is optically thick, self-shielding cannot raise the CO lifetime to the $\sim 1$\,Myr
required for this to be a viable solution.
Moreover, since photodissociation is driven by the interstellar radiation field, the problem cannot be 
circumvented by devising a thin disk geometry to prevent penetration of the stellar radiation. 
The solution may be to invoke an unseen gaseous component, such as hydrogen which would be abundant if
the gas is primordial.
However, the resulting collisions would imply that the 5.6\,K excitation temperature inferred from CO line
ratios is the kinetic temperature of the disk (i.e., that the disk is in LTE), which is significantly
lower than that expected from protoplanetary disk observations.
\citet{Hughes2017} reached a similar conclusion for the 49 Ceti gas disk, that the temperature and gas scale
height require a relatively high mean molecular weight, disfavouring a primordial origin for the gas
in this disk.

Thus HD21997 presents a puzzle, and similar considerations apply to
other young A stars with bright debris disks and significant levels of CO gas emission
\citep[e.g.,][]{Pericaud2017}.
Nevertheless, it is worth noting that while roughly half of young A stars with bright debris disks have gas,
it is only $\sim 10$\% of young A stars that have bright debris disks \citep{Carpenter2009},
which means that not all A stars necessarily go through this phase.
% Lieman-Sifry sample requires F70/F*>100 which from Fig. 6 of Wyatt 2008 means <20%, presumably similar for the f=0.05-1% criterion
Moreover, there is no evidence yet that this is not a phenomenon unique to A stars;
e.g., \citet{Kral2017} devised a simple prescription to predict the expected level of secondary CO, and
so quantify whether any of the current debris disk gas detections look anomalous, only identifying
two further A stars HD131835 and HD138813 (also implying that most of the detections in Fig.~\ref{fig:1} could be
consistent with secondary gas production).
On the other hand, the persistence of primordial gas in some systems needs to be understood, since this
could apply to other systems albeit at lower gas levels.

%%%%%%%%%%%%%%%%%%%%%%%%%%%%%%%%%%%%%%%%%%%%%%%%%%%%%%%%%%%%%%
\subsubsection{Implications of debris disk gas}
Within the context of this review, a few things can be concluded.
{\bf (i)}
The dispersal of primordial gas may be inefficient and leave significant quantities for 10s of Myr.
This leaves open the possibility that gas is present that can be accreted onto planetary cores and so
lead to continued growth of planetary atmospheres.
If sufficient gas is present, this can also affect the dynamical evolution of the planetary system,
for example damping planetary eccentricities and inclinations and leading to planet migration (see chapter by Morbidelli).
Even moderate quantities of gas can affect the evolution of small dust grains, which could
affect the interpretation of the observed dust properties \citep[e.g.,][]{Takeuchi2001,Lyra2013}.
For example, the presence of gas may prevent the removal of micron-sized dust that would otherwise be expelled by
radiation pressure, thus explaining the anomalously high dust temperature (given its radial location)
for some hybrid disks \citep{Lieman-Sifry2016,Moor2017}.
{\bf (ii)}
Some fraction of the CO that is present in protoplanetary disks ends up in planetesimals
later to be released.
It is unclear whether the quantity that is sequestered from the disk is sufficient to affect the
observed chemical structure of protoplanetary disks (see chapter by Bergin et al.).
Nevertheless, secondary debris disk gas may be observational evidence for a process that started in
the protoplanetary disk, thus allowing the two evolutionary phases to be connected.
Moreover, debris disk gas observations already provide evidence for the volatile-rich composition of
extrasolar planetesimals in some systems.
This allows consideration of whether such volatiles can later be delivered to
planets in the inner regions of the system, thus aiding conditions conducive to the development of life.
The detailed composition of debris disk gas, e.g. from the detection of other molecular tracers like CN
\citep[e.g.,][]{Matra2018},
would also inform on processes in the protoplanetary disk such as condensation sequences (see chapter by Pudritz et al.).

%%%%%%%%%%%%%%%%%%%%%%%%%%%%%%%%%%%%%%
\section{Probing the outcome of planetesimal formation}
\label{s:pll}
Our understanding of dust evolution in protoplanetary disks has recently undergone significant advances
(see chapter by Andrews \& Birnstiel).
For example, observations with ALMA provide observational evidence for dust growth and drift in broad
agreement with theoretical expectations for its interaction with the gas disk, at least once the
gas disk is allowed to have complex structure.
However, there is as yet no observational evidence that planetesimals are present during the protoplanetary
disk phase.
Indeed such evidence would be hard to come by, because observations are only sensitive to
dust smaller than $\sim 1$\,cm in size.
Planetesimals (i.e., $>1$\,km objects) may be present, but their low opacity means that for realistic
masses these would have no observable signature.
While these same planetesimals may collide and produce dust that can have an observable signature,
given the possible presence of gas to entrain the dust and the young age of systems with protoplanetary
disks (up to a few Myr), it would be very hard to argue that the dust could not be primordial.

%%%%%%%%%%%%%%%%%%%%%%%%%%%%%%%%%%%%%%%%%%%%%%%%%%%%%%%%%%%%%%
\subsection{Baseline model of debris disks: planetesimal belt}
This same argument does not apply to dust seen in the debris disks around $>10$\,Myr main sequence stars, for
which the short lifetime of the observed dust is used to infer that larger planetesimals must be present.
Such large bodies can survive over the age of the star in the face of collisions and radiation
forces, providing a source population that continually replenishes the observed dust.
This picture of debris disk dust being replenished in a steady state manner from collisions amongst
belts of planetesimals has gained widespread support due to:
images showing that mm-sized grains in many systems are confined to narrow rings
\citep[e.g., see Fig.~\ref{fig:4},][]{Marino2016,MacGregor2017}, 
the inferred size distribution of mm-sized grains fitting with expectations of a collisional cascade
\citep{MacGregor2016},
the presence of a halo of micron-sized grains outside these rings as expected due to radiation pressure
\citep{Strubbe2006},
the size distribution of small dust close to the radiation pressure blow-out limit agreeing
broadly with expectations \citep{Pawellek2014},
the manner in which disks are fainter around older stars being consistent with their depletion in
collisional erosion assuming the presence of planetesimals at least a few km in size
\citep{Wyatt2007,Lohne2008,Gaspar2013}.

This means that debris disks provide valuable information on the outcome of planetesimal formation processes,
showing where such planetesimals formed (or rather where they ended up), the size of those planetesimals,
the total mass contained in the belts, and the level of stirring (i.e., collision velocities
between planetesimals).
However, it is often the case that this information is not uniquely constrained even for well studied disks.
For example, for Fomalhaut for which imaging finds a belt radius of $\sim 130$\,au and
constrains the level of stirring (see chapter by Kalas), it is only possible to say that given its 440\,Myr age, the planetesimals must be
larger than a few km in size requiring a total mass of at least 10s of $M_\oplus$ \citep{Wyatt2002}.
For most systems however, the radial location of planetesimals is much less well constrained and must be
estimated from the dust temperature, and conclusions are degenerate with assumptions about the stirring level
(and about planetesimal strength). 
Nevertheless, planetesimal sizes of 10s of km and disk masses of $>1M_\oplus$ are the right ballpark for
debris disks that can be detected.

\begin{figure}
%\vspace{-0.8in}
%\hspace{-0.9in}
\includegraphics[scale=0.52]{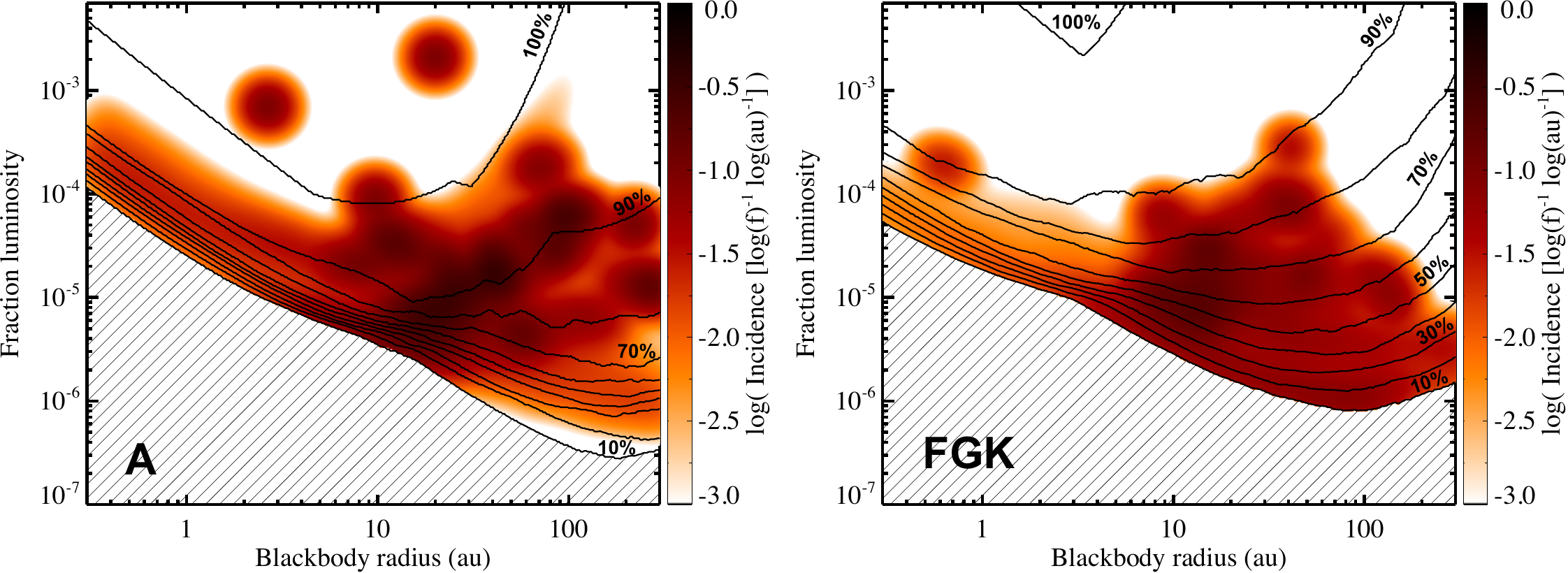}
%\vspace{-1.5in}
\caption{Fraction of stars with debris disks of given fractional luminosity and black body radius, for the nearest
100 A stars (left) and the nearest 300 FGK stars (right) \citep[using the samples in][]{Phillips2010}.
The true disk radius is likely to be a factor of a few larger than the black body radius because the small
grains that dominate the emission have relatively high temperatures due to their inefficient emission
\citep{Booth2013}.
The contours show the fraction of stars in these samples for which disks could be detected, going from 10\% to 100\%
in 10\% increments.
This detection bias has been corrected for and disk incidence only plotted in regions where disks could be
detected for $>10$\% of the stars, the remaining area being shown with cross-hatching.
(Figure made by G. Kennedy using the technique described in Sibthorpe et al.).}
\label{fig:2}       % Give a unique label
\end{figure}

%%%%%%%%%%%%%%%%%%%%%%%%%%%%%%%%%%%%%%%%%%%%%%%%%%%%%%%%%%%%%%
\subsection{Population model fits to debris disk statistics}
Population models \citep[e.g.,][]{Wyatt2007,Lohne2008,Gaspar2013,Krivov2018}
make the assumption that all disks have the same level of stirring, planetesimal strength
and maximum planetesimal size, and use that to derive the underlying distribution of disk radii and initial masses, since
that then sets the observable properties of disks expected to be found around a star of given age and spectral type
which can be compared with observations.
Fig.~\ref{fig:2} summarises our current understanding of the observed disk properties, with the colour scale
showing the fraction of stars that have a disk of given fractional luminosity and radius for the $\sim 300$ nearest FGK stars (left) and
$\sim 100$ nearest A stars (right).
These plots combine information from sufficient disks that the distributions are relatively smooth, although the blobs
in the sparsely populated region of high fractional luminosity disks result from individual disks.
For both spectral types, approximately 20\% of stars have detectable disks \citep{Eiroa2013,Thureau2014,Sibthorpe2017},
and with no far-IR space mission imminent, the complement of (cold) debris disks around nearby stars is unlikely to
increase significantly any time soon.
There are several well understood biases in detectability in this figure, which are explained in \citet{Wyatt2008},
but their implications can be understood from the contours on Fig.~\ref{fig:2} which show the fraction of stars in the sample
for which the observations would have been able to detect a disk in this part of parameter space
(e.g., we cannot know what fraction of stars have disks that lie in the cross-hatched region).

The populations for both spectral type groupings are similar, in that they show disks concentrated at radii
from 10-100\,au, with fractional luminosities $10^{-6}-10^{-4}$.
They also both exhibit an upside-down V-shape for the upper envelope where disks are found in Fig.~\ref{fig:2}.
This shape is in good agreement with the predictions of steady state collisional population models.
In such models the small radius side of this upper envelope ($<30$\,au) is caused by collisional depletion
over the stellar age, since steady state collisional erosion is predicted to cause disks of the same radius
to have evolved by a given age to the same mass and luminosity that is independent of their initial mass
(a disk that is initially more massive will start out brighter but
decay faster than one that is less massive, to end up at the same level).
Since the disks included in Fig.~\ref{fig:2} should all have suffered approximately Gyr of evolution, all close-in
disks should lie below the same envelope, and any that have dust luminosities that lie significantly above this 
are inferred to be a transient phenomenon, since they cannot be explained by steady
state processes \citep{Wyatt2007b}, unless they are found around young ($\ll 100$\,Myr) stars.
The large radius side of the upper envelope in the population models is set by the maximum initial mass present in the belts,
and so such large disks ($\gg 30$\,au) are inferred to have yet to undergo significant collisional depletion
\citep[due to the long collision timescales at this distance, see discussion in][]{Wyatt2007}.
For now, such population models have been used to show that steady state collisional erosion
can explain most of the observed trends for A stars \citep{Wyatt2007}, FGK stars \citep{Lohne2008,Kains2011}
and M stars \citep{Morey2014} \citep[see also][]{Gaspar2013}, and thus also provide a plausible population from which
to consider the detectability of such disks.
However, it should be noted that any conclusions about the debris disks of the 80\% of stars without detected disks
rely on extrapolations from the known disks and include assumptions about the form of
the underlying mass and radius distributions of the planetesimal belts.
Ultimately though, it will be possible to use these population models to set constraints on
the distribution of planetesimal belt masses and radii that emerge from the protoplanetary disk phase.

Comparing the observed distributions around A and FGK stars in Fig.~\ref{fig:2}, it is evident that the
peak in the upside-down V (of the upper envelope of the dense coloured region where most disks
are found) is at larger black body radius for the more massive stars.
This implies that collisional erosion has progressed to larger distances around A stars,
despite these stars being on average younger than the FGK stars.
This could mean that collisional evolution is faster around the earlier spectral types,
perhaps because the mass in their disks is dominated by smaller planetesimals.
However, such an inference is complicated by the fact that the population of FGKs stars
would be expected to shift further to the right if plotted against the true disk
radius (rather than that inferred from the dust temperature assuming black body grains). 
Nevertheless, it seems that the debris disks of A stars and FGK stars do have some
intrinsic differences (although the exact spectral type boundary at which any difference
occurs is poorly constrained);
e.g., it has also been suggested that A star disks tend to be more massive than those of
FGK stars \citep{Greaves2003}.

%%%%%%%%%%%%%%%%%%%%%%%%%%%%%%%%%%%%%%%%%%%%%%%%%%%%%%%%%%%%%%
\subsubsection{Beyond the baseline model}
\begin{figure}
%\vspace{-0.8in}
%\hspace{-0.9in}
\includegraphics[scale=0.55]{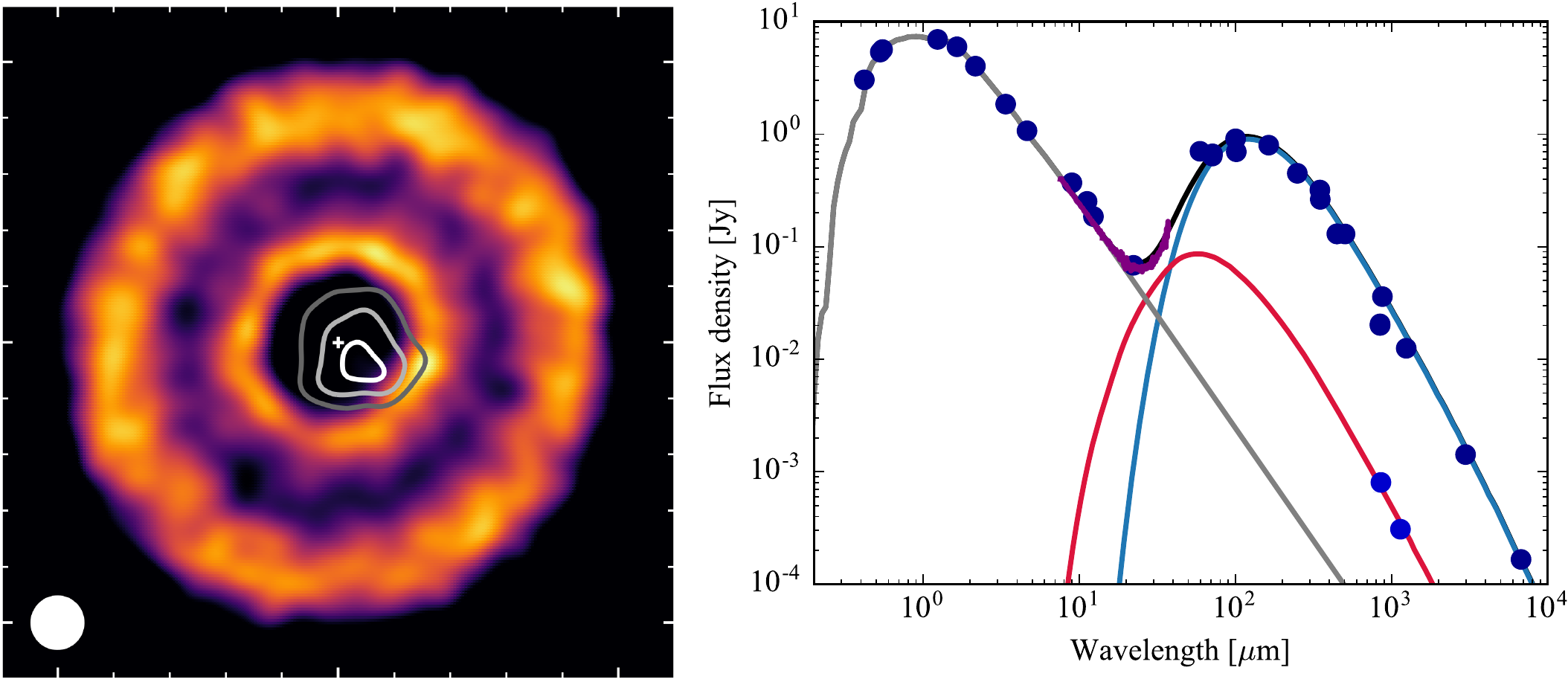}
%\vspace{-1.5in}
\caption{Observations of the debris disk of the 80-200\,Myr G2V star HD107146 (Marino et al., in prep.).
(Left) ALMA images of the dust emission show that the broad (30-150\,au) cold debris disk is resolved
into two rings with a partially filled gap centred at 80\,au.
Emission from an additional warm dust component at 10\,au is also evident in the residual contours
(the emission remaining after subtraction of a model for the cold component).
This warm component is also seen as excess mid-infrared emission in the spectral energy distribution
of the star (right), which requires a warm temperature component (red) in addition to the emission from the 
star (grey) and the cold component (blue).
}
\label{fig:3}       % Give a unique label
\end{figure}

Detailed observations of individual disks also show ways in which the population models can be improved.
For example, not all disks have planetesimals concentrated in narrow rings.
Several are shown to be radially very broad \citep{Booth2013}, extending over a factor of 2 or more in radius.
Collisional evolution of such disks can lead to a flat surface brightness profile \citep{Schuppler2016} which can
be very difficult to resolve in the sub-mm even when the disk is bright in the far-IR \citep{Marino2017b}.
Other disks have been found to have dust concentrated at multiple radial locations.
This was first inferred from the infrared spectrum which implied two temperature components were
needed to fit the spectrum \citep[][see e.g. Fig.~\ref{fig:3} right]{Chen2014}, requiring hot dust significantly closer to these
stars than their cold outer belts \citep{Su2013,Kennedy2014}.
There are also examples of disks for which high resolution imaging has shown a broad outer belt
to be comprised of two rings with a gap inbetween \citep[see Fig.~\ref{fig:3} left,][]{Ricci2015,Golimowski2011}, or in which
a narrow ring has been found to have a fainter ring just outside \citep{Marino2016}.
Such multiple components would be very hard to discern from the spectrum, and could
indicate locations of preferential planetesimal formation (e.g., related to the ring-like structures
seen in protoplanetary disks, see chapter by Andrews \& Birnstiel), or provide evidence for sculpting by planets
(see later discussion).

High resolution imaging also allows the level of stirring in the belt to be estimated. 
For example, through the scale height of edge-on disks (which gives direct information on dust particle
inclinations), or through the detailed structure of the halo of small dust extending beyond the ring
\citep[since small halo dust would be underabundant in a disk with low stirring as it is radiation forces
not stirring that sets its depletion rate,][]{Thebault2008}.
While for most debris disks the origin and level of stirring is unknown,
it may be notable that some disks have been inferred to have very low levels of stirring (eccentricities $<1$\%),
below which collisional depletion of the belts becomes very slow
and the collisional production of small grains is prohibited rendering such disks relatively 
hard to detect \citep{Krijt2014}.
Indeed, the existence of unstirred disks (eccentricities $\ll1$\%) of 10-100\,m planetesimals that
can survive Gyr of evolution has been proposed to explain the detection of faint anomalously
cold debris disks with Herschel \citep{Krivov2013}, and so it may be that many unstirred debris disks
lurk below the detection threshold on Fig.~\ref{fig:2} \citep{Heng2010}.
That the absence of detectable emission does not necessarily indicate an absence of planetesimals
should be evident from the fact that the Solar System's Kuiper belt would not have been detected
in any of the current far-IR surveys of nearby stars \citep{Vitense2012}.

Given the difficulty of detecting planetesimals in protoplanetary disks mentioned above, there is
for now very little information about when planetesimals formed.
It is generally assumed that planetesimals form relatively early on ($\ll 1$\,Myr) so that they can provide the
building blocks for planets that are also thought to form before the protoplanetary disk disperses.
Evidence from the Solar System also supports an early formation epoch for the asteroids \citep{Wadhwa2007}.
However, it is typically the case that $>1M_\oplus$ of dust is present in protoplanetary disks
up to the epoch of its dispersal (see Fig.~\ref{fig:1}), leading to the possibility that at least one population
of planetesimals could form during disk dispersal \citep[e.g.,][]{Carrera2017,Ercolano2017}.
Observations of very young populations of debris disks (such as those around the class III stars found in
star forming regions, see Fig.~\ref{fig:1}) will help to disentangle these early phases.

%%%%%%%%%%%%%%%%%%%%%%%%%%%%%%%%%%%%%%
\section{Debris disks as a component of a planetary system}
\label{s:pl}
While debris disks provide evidence for the successful formation of planetesimals
in the outer regions (typically 10s of au) of many (at least 20\%) protoplanetary disks,
this does not automatically indicate that planets also formed in these systems.
Similarity of the observed disks to the Solar System's Kuiper belt \citep{Currie2015,Booth2017}
provides circumstantial support for the presence of a planetary system, but hard evidence is
needed to support this anthropocentric view.
Such evidence can be found for the population as a whole by comparing samples of stars with debris disks and/or
exoplanets, and for some individual systems from the detailed structure of their disks.

%%%%%%%%%%%%%%%%%%%%%%%%%%%%%%%%%%%%%%%
\subsection{Correlations of debris disks and exoplanets}
Evidence for a difference between the far-IR properties of the debris
disks around stars with and without planets was lacking from studies of Spitzer data \citep{Bryden2009},
but has now been found following more recent surveys with Herschel that were both more sensitive
and covered a larger number of exoplanet host stars \citep[see][]{Matthews2014}.
The disks appear to be on average slightly brighter when radial velocity planets are present.
There is also evidence for a dependence on planet mass, in that systems with planets but where none is
more massive than Saturn tend to have a higher incidence of detectable debris \citep{Wyatt2012}.
However, such correlations remain tentative \citep{MoroMartin2015}.
At such vastly different spatial scales the planets ($\ll 5$\,au) are unlikely to have any direct influence on the disk
($\gg 20$\,au) detectability.
Thus, if confirmed, these correlations could either point to the planets having an indirect influence on the outer
planetesimals \citep[e.g., because systems of high mass planetary systems scatter smaller planets out that can deplete
an outer disk,][]{Raymond2011}, or to the outcome of planetesimal and planet formation at
the different locations being linked through initial protoplanetary disk properties.
For example, the increased debris disk brightness for the planet-host stars is at a level
compatible with that expected if such systems formed from protoplanetary disks that are amongst the top 6\%
of the population in terms of their solid disk masses \citep{Wyatt2007c}, which is supported by the
planet-metallicity correlation \citep{Fischer2005} and possibly
also by a correlation of debris disk mass with metallicity \citep{Gaspar2016}.
Identifying a correlation of debris disks with planets imaged in the outer regions of the
systems would be particularly informative, and it is notable that all currently known imaged planets are in systems
with debris disks \citep{Bowler2016}.

%%%%%%%%%%%%%%%%%%%%%%%%%%%%%%%%%%%%%%
\subsection{Structures associated with planets}
Disk structure can be a very sensitive tracer of planets because the orbital motion of
planetesimals is inevitably perturbed by the presence of planets. 
The same is true in protoplanetary disks, but the presence of significant quantities of gas
makes linking observed dust structures to planets more complicated (see chapter by Andrews \& Birnstiel).
There is also complexity when interpreting debris disk dust observations, since the dust is
affected by radiation forces.
However, the way collisions and radiation forces cause the radial distribution of dust to differ
from that of the planetesimals is reasonably well understood \citep{Krivov2006}, and through
a range of modelling efforts it is also possible to understand how the two distributions
are related even in more complex dynamical situations resulting in non-axisymmetric structure
\citep{Wyatt2006}.
Thus interpretation of debris disk structures starts with consideration of how planetesimal
orbits are affected by a planet's gravity.
Such perturbations come in three types: secular, resonant and scattering.

%%%%%%%%%%%%%%%%%%%%%%%%%%%%%%%%%%%%%%
\subsubsection{Secular structures}
Secular perturbations are long range interactions and allow a planet to modify the orbit of planetesimals
even at large distance.
For planets on relatively low eccentricity orbits ($e \ll 0.3$), such perturbations
cause planetesimal orbits to become eccentric, with pericentres that precess at rates that
depend on location in the disk.
This can cause initially coplanar circular planetesimal orbits to cross, resulting in collisional
destruction \citep{Mustill2009}, and setting up a tightly wound spiral wave that propagates
through the disk \citep{Wyatt2005}, eventually causing a disk to become eccentric \citep{Wyatt1999}.
If the planet is inclined to the disk, the planetesimals' orbital planes also precess, again
at a rate dependent on location, which can cause the disk to appear warped when viewed edge-on
\citep{Augereau2001}.
The secular perturbations of a highly eccentric planet, such as one scattered out by an inner
planetary system, can cause bell-shaped structures, disks that are orthogonal to the planet's orbit,
and double-ringed structures \citep{Pearce2014,Pearce2015}.

\begin{figure}
%\vspace{-0.8in}
%\hspace{-0.9in}
\includegraphics[scale=0.68]{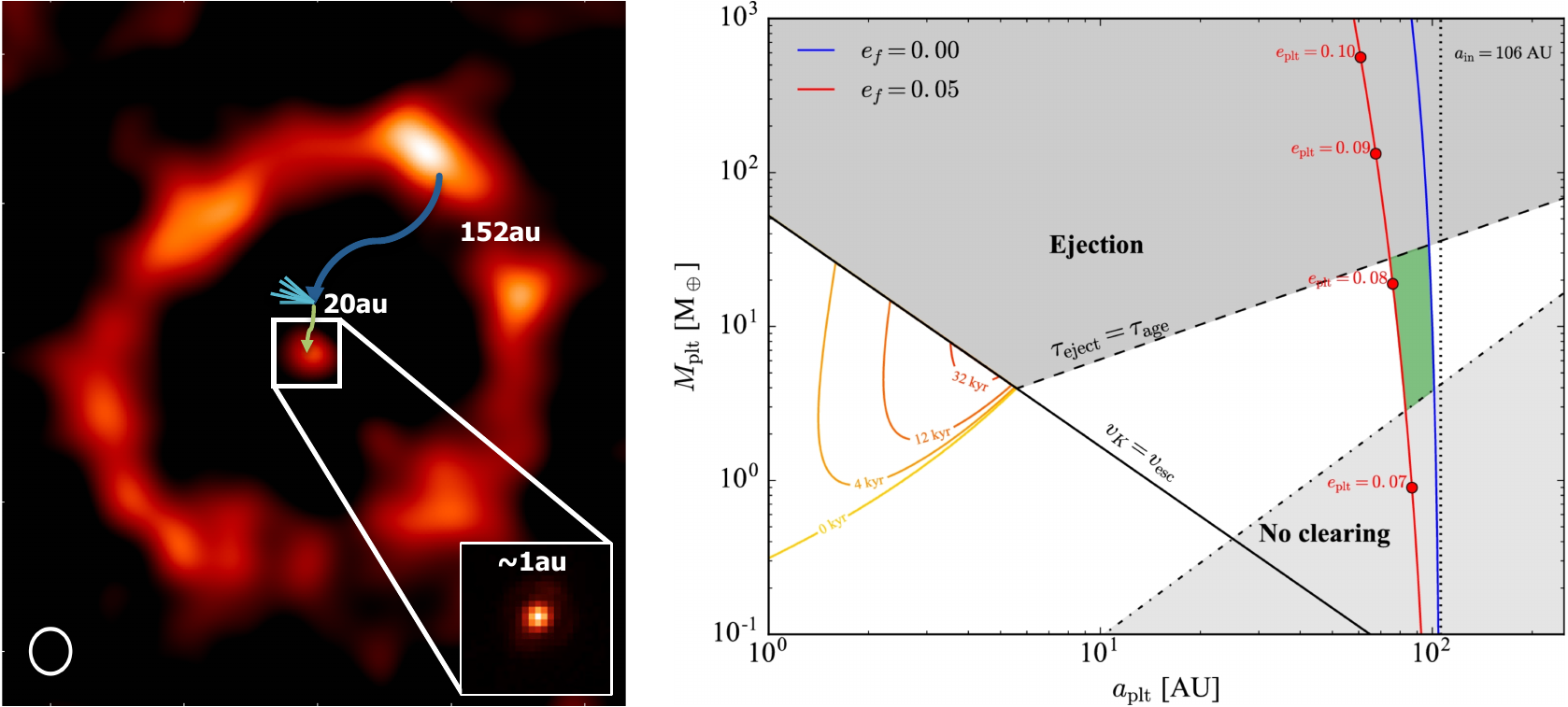}
%\vspace{-1.5in}
\caption{Observations of the debris disk of the 1Gyr F2V star $\eta$ Corvi (left) and
implications for its underlying planetary system \citep[right,][]{Marino2017}.
The debris disk has three components: a cold outer belt at 152\,au imaged with ALMA (main image),
CO gas at $\sim 20$\,au thought to originate in
icy planetesimals sublimating as they are scattered into the inner system (cartoon depiction),
and an inner hot component at $\sim 1$\,au possibily the refractory component of those
planetesimals or the product of a collision with a planet \citep[inset image][]{Smith2008}.
If there is a planet sculpting the disk's inner edge this must lie in the green shaded
region with limited eccentricity due to the circular appearance of the disk.
Inward scattering of planetesimals is inhibited if any planets lie in the {\it ejection}
region \citep{Wyatt2017}.
The yellow and orange lines show how long collisional debris from such planets would remain
detectable.
}
\label{fig:4}       % Give a unique label
\end{figure}

While most of these predicted features have been observed in debris disk images, the strongest evidence
that these originate in planetary perturbations comes from a warp in the $\beta$ Pictoris disk, for which
the warp was used to predict the existence of a $9M_{\rm jup}$ planet at 9\,au that was later discovered
by direct imaging \citep{Lagrange2010}.
On the other hand, the structures associated with eccentric debris rings seem to be ubiquitous,
possibly explaining large scale asymmetric features of scattered light images
\citep[such as those called {\it needles} and {\it moths},][]{Lee2016,Lohne2017},
as well as a brightness asymmetry in the infrared that undergoes a $180^\circ$ phase shift depending on wavelength
\citep[i.e., switching from {\it pericentre glow} to {\it apocentre glow},][]{Pan2016}.
The lack of eccentricity of a debris ring also limits the eccentricity of any orbiting
companions \citep[see Fig.~\ref{fig:4},][]{Marino2017}.
The best characterised eccentric debris ring is that of Fomalhaut \citep[see chapter
by Kalas,][]{Kalas2013,MacGregor2017},
but the way in which this informs on the underlying planetary system is complicated by the discovery of
a planet-like object Fomalhaut-b which orbits near the belt but cannot be responsible for
the ring eccentricity \citep{Tamayo2014,Beust2014}.
It should be noted that the long range nature of secular perturbations means that perturbing objects could
be either interior or exterior to the debris belt \citep{Nesvold2017}, and indeed it is suggested that the
eccentricity in this system could arise from external perturbations from the companion
stars \citep{Shannon2014,Kaib2017}.

%%%%%%%%%%%%%%%%%%%%%%%%%%%%%%%%%%%%%%
\subsubsection{Resonances: gaps, clearing and clumps}
A planet's resonant perturbations only apply to planetesimals with orbital periods within narrow ranges.
Nevertheless, resonant planetesimals can be overabundant due to resonant trapping during planet migration.
The geometry of resonant orbits makes the distribution of planetesimals clumpy,
allowing observed structures to reveal the mass, current location and past migration history
of the perturbing planet \citep{Wyatt2003}.
The best evidence for resonant structures in debris disks is the clump in the $\beta$ Pictoris disk
\citep{Telesco2005,Dent2014,Matra2017}, with tentative signs in other disks
\citep{Hughes2012,Booth2017}.

Planetary resonances also cause the region near a planet to be cleared due to chaos from overlapping resonances
\citep{Wisdom1980}.
Thus debris disk edges set constraints on the possible perturbing planet's mass and location
(see Fig.~\ref{fig:4}). 
Their detailed shape can also be a powerful indicator of the mass of the
planet sculpting the disk;
lower mass planets cause sharper edges \citep{Chiang2009}, although this shape 
also depends on planetesimal (and planet) eccentricity \citep{Mustill2012}.
The existence of planetary systems inside debris rings would
be a natural explanation for the dearth of planetesimals there (see Fig.~\ref{fig:4}), requiring typically around 5 Neptune mass
planets \citep{Faber2007}.
Planets may also be responsible for carving wide gaps in broad debris disks, with gap sizes 
indicative of the total number and masses of planets present \citep{Nesvold2015,Shannon2016},
although broad gaps can also be carved by secular resonances that sweep through the disk
as the protoplanetary disk disperses \citep{Zheng2017}.
Planets can also carve very narrow gaps in debris disks at specific mean motion resonances \citep{Tabeshian2016},

If planets are not responsible for shaping the inner edges of debris disks and for the absence of
planetesimals in the inner regions, then another explanation for these features must be sought.
Collisional depletion of the inner regions could explain a shallow inner edge at a location
determined by the system age and planetesimal size \citep{Kennedy2010,Schuppler2016}.
A preferential location for planetesimal formation, say at a snowline or dead zone edge in a protoplanetary
disk, could also explain an overdensity at a certain location. 
However, since the location of such transitions in protoplanetary disks evolve with time \citep{Hasegawa2012,Eistrup2017},
it remains to be seen if these could explain a sharp disk edge for which an element of shepherding is needed.
The presence of gas in debris disks has also been proposed as a mechanism to form sharp edges
\citep{Lyra2013,Richert2017},
although it is unclear that sufficient quantities of gas are present.

%%%%%%%%%%%%%%%%%%%%%%%%%%%%%%%%%%%%%%
\subsubsection{Scattering: exocomets and scattered disks}
It might be expected that scattering of planetesimals by a planetary system is only relevant in the
earliest phases of evolution, because objects that can have close encounters with planets are short-lived
\citep[e.g., comets leaving the Kuiper belt in the Solar System have a dynamical lifetime of $\sim 45$\,Myr,][]{Levison1997}.
However, for low mass planets in the outer regions of disks scattering timescales can be longer than
the system age, allowing a long-lived scattered disk of primordial planetesimals and the possibility
that planet-mass bodies could reside within debris disks \citep{Wyatt2017}.
This could result in a continual injection of planetesimals into a planetary system \citep[e.g.,][]{MunozGutierrez2015},
thus replenishing a population of exocomets that need not be massive to be detectable.

There is a growing body of observational evidence for hot dust in the inner regions of planetary systems,
with debate ongoing as to whether this originates in in-situ asteroid belts
\citep[which is possible if the belts are beyond at few au,][]{Wyatt2007b,Geiler2017},
dust dragged in from the outer belt by Poynting-Robertson drag
\citep[which is a plausible explanation for the low 0.1-1\% excess levels of hot dust seen by KIN,][]{Wyatt2005b,Mennesson2014},
recent giant impacts between planetary embryos
\citep[which is the preferred explanation for extreme $\gg 10$\% levels of hot dust that are found
predominantly around $<100$\,Myr stars,][]{Wyatt2016},
or the destruction of exocomets scattered in from an outer belt
\citep[which is supported by the presence of CO gas in one system with moderate $\sim 10$\% excess levels,
see Fig.~\ref{fig:4},][]{Marino2017}.
Exocomets passing in front of the star are also the preferred explanation for the variable gas absorption seen
toward $\beta$ Pictoris \citep{Beust1996,Kiefer2014}, and for dips in the light curves of some main sequence
stars \citep{Boyajian2016,Rappaport2017}.

Exocometary activity may be expected at some level in all systems in which a planetary system interacts with a planetesimal
belt, e.g., through either exterior or interior resonances \citep{Faramaz2017}.
However, for some systems exocometary scattering may be more efficient, and it is such systems that are likely
to have the highest levels of observed exocometary dust (e.g., like $\eta$ Corvi, Fig.~\ref{fig:4}).
The efficiency with which a planetary system passes exocomets in from an outer belt has been studied both numerically
and analytically, leading to the conclusion that closely packed systems of low mass planets are the most efficient
\citep{Bonsor2012}, and highlighting the importance of the lack of massive planet that is an efficient
ejector \citep[see Fig.~\ref{fig:4},][]{Wyatt2017}.
While this might suggest that the super-Earth planetary systems discovered by {\it Kepler} are efficient exocomet
scatterers, the planet chain must also extend out as far as a long-lived source population of comets, which
as noted earlier must likely reside $>30$\,au to survive for Gyr against collisional depletion.
Young planetary systems may host an exocomet population that is a residual of their formation process
(which hence could potentially be a valuable probe of that process), but any such
exocomet population will decrease over time as planetesimals are removed by scattering.
Ways to maintain an exocomet population over Gyr timescales have also been sought, e.g., by allowing a planet to migrate
into a disk \citep{Bonsor2014} or by witnessing the aftermath of a dynamical instability \citep{Bonsor2013}.
For now the interpretation of hot debris has many uncertainties, but since this dust (and the planetesimals from 
which it originates) must pass through the planetary system, the architecture of those planets is inevitably
imprinted on both the level, radial distribution and asymmetrical structure of the dust.

%%%%%%%%%%%%%%%%%%%%%%%%%%%%%%%%%%%%%%
\section{Witnessing ongoing planet formation}
\label{s:ongoing}
Planet formation processes are thought to take place predominantly during the protoplanetary
disk phase, since that is when the circumstellar disk contains sufficient mass in both dust and
gas to form planets (see Fig.~\ref{fig:1}).
By the debris disk phase, observations show that any significant solid mass must be either in
planetesimals or planets, and the gas content is also believed to be relatively limited (but see above).
Nevertheless, the processes associated with planet formation can continue in the absence of a protoplanetary
disk.
For example, planetesimals may continue to grow into dwarf planets, and planets may also continue to grow
through collisions or accretion of gas. 
Similarly, the planetary system that emerges from the protoplanetary disk may undergo significant evolution
(e.g., through dynamical instability or planet migration).
Evidence for all of these processes may be imprinted on the debris disk.

%%%%%%%%%%%%%%%%%%%%%%%%%%%%%%%%%%%%%%%%%%%%%%%%%%%%%%%%%%%%%%
\subsection{Continued growth of dwarf planets}
What stirs a debris disk is unknown, that is, why the planetesimals collide at sufficient velocity to
break apart.
It might be thought more likely that planetesimals formed in
a low collision velocity environment, since this would aid their growth,
and regardless of their formation mechanism, gas drag would act to damp collision
velocities while in the protoplanetary disk.
Perhaps the neatest explanation is that a debris disk stirs itself.
In a series of papers, \citet{Kenyon2002,Kenyon2008,Kenyon2010} showed how a disk containing only planetesimals colliding
at low velocity would eventually grow Pluto-sized objects that stir the disk in its vicinity causing 
a collisional cascade and the ultimate destruction of the debris disk.
Since growth timescales are longer further from the star, this leads to the prediction that a broad
planetesimal disk would appear as a bright ring of emission that propagates out to the edge of the disk over
Gyr timescales.

There is, however, no evidence to corroborate this picture.
For example, observations of how disk brightness decreases with stellar age show that, on a population
level, the observations are consistent with stars being born with narrow debris belts that
are pre-stirred by the time the protoplanetary disk disperses \citep{Kennedy2010}.
If there is any evidence for disks having larger radii around older stars, it is consistent with
the smaller radii disks being collisionally depleted below the detection threshold, and inconsistent with
seeing different parts of a broad disk at different epochs. 
Nevertheless, these studies show that the growth of dwarf planets would occur naturally
if protoplanetary disks leave behind unstirred planetesimal disks.
This picture is also consistent with the observed debris disk populations (e.g., Fig.~\ref{fig:2})
as long as the planetesimals are usually confined to narrow rings,
although it would still be necessary to invoke some other stirring agent to
explain the very young $<20$\,Myr debris disks seen at 10s of au, which seem to
be stirred on much shorter timescales than permitted by the model \citep{Milli2017}.

%%%%%%%%%%%%%%%%%%%%%%%%%%%%%%%%%%%%%%%%%%%%%%%%%%%%%%%%%%%%%%
\subsection{Giant impacts at 10s of au}
Disk structure provides an alternative test for the possible growth of dwarf planets within debris
disks. 
Such planets grow through multiple collisions between dwarf planet embryos that would be expected
to release significant quantities of dust that should be readily detectable.
While the resulting cloud of unbound debris would appear as an expanding clump on short timescales that
is unlikely to be detected \citep{Wyatt2002}, the resulting structure would be asymmetric for many
1000s of orbits due to the passage of the debris through the point in space at which the impact occurred
\citep{Jackson2014,Kral2015}.
This results in a significantly increased collision rate at this point (and is thus where dust and gas
is produced most vigorously) and translates into detectable asymmetric structures for Myr for collisions in the
outer regions of the disk.

This was suggested as the origin of the dust and gas clump in the $\beta$ Pictoris disk \citep{Dent2014}, although
the tentative orbital motion of the clump \citep{Li2012} and the broad radial structure of the clump
\citep{Matra2017} now disfavours this explanation.
Giant impacts in the outer regions of debris disks have also been proposed as the origin of
structures in the HD181327 and HD61005 disks \citep{Stark2014,Olofsson2016}.
Such structures are inevitable at some level and frequency if large planetesimals and
dwarf planets are present in debris disks.
Thus if these features are not observed this would be informative about the maximum
number of dwarf planets that could be present.

\begin{figure}
%\vspace{-0.8in}
%\hspace{-0.9in}
\includegraphics[scale=0.51]{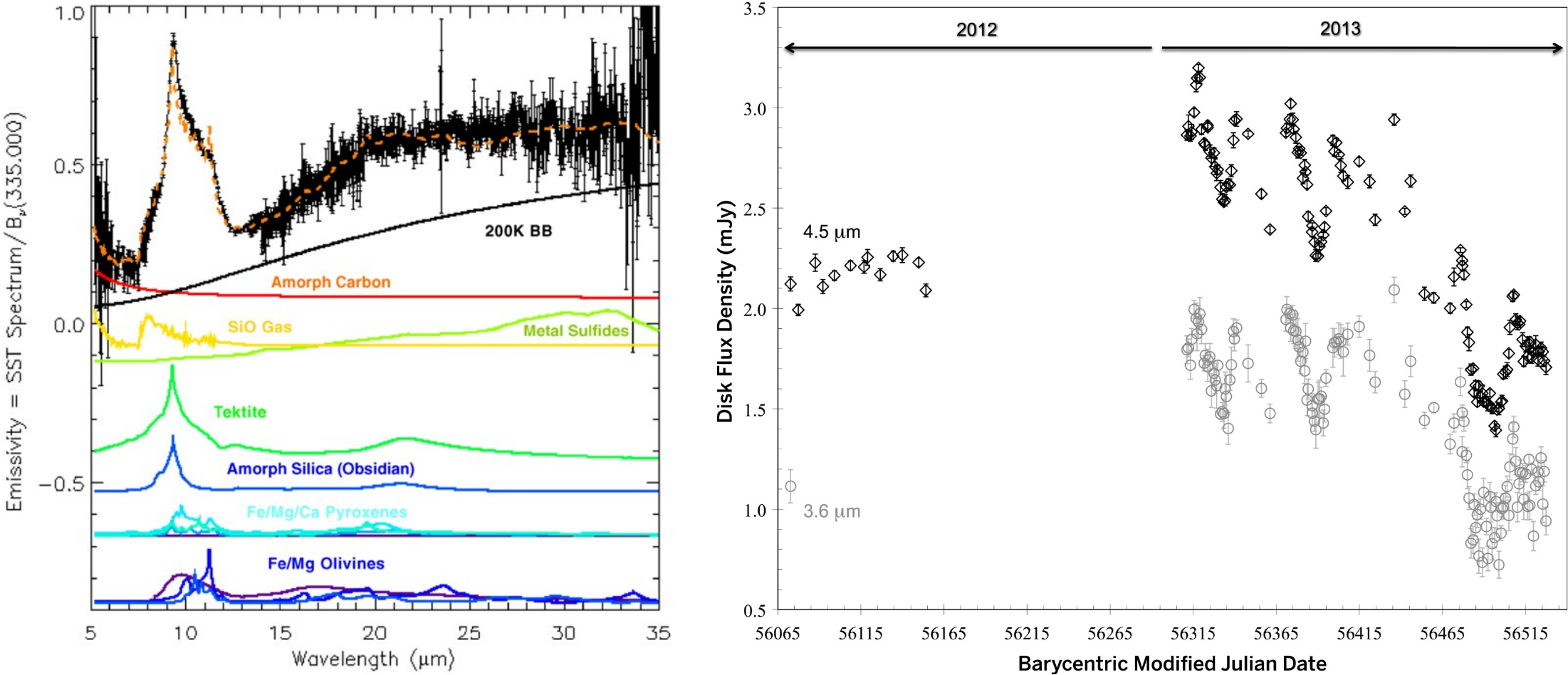}
%\vspace{-1.5in}
\caption{Spectral information \citep[left,][]{Lisse2009} and temporal evolution \citep[right,][]{Meng2014}
of the infrared emission from some young stars provides the strongest evidence for giant impacts in the
terrestrial planet region.
The spectrum is that of the 22\,Myr A3V star HD172555 for which modelling shows evidence for the
amorphous silica expected to be produced in a high velocity collision.
The rapid cyclical evolution in brightness of the 35\,Myr G6V star ID8 is consistent with that expected from
an optically thick clump of debris produced in a collision.
}
\label{fig:5}       % Give a unique label
\end{figure}

%%%%%%%%%%%%%%%%%%%%%%%%%%%%%%%%%%%%%%%%%%%%%%%%%%%%%%%%%%%%%%
\subsection{Giant impacts in the terrestrial planet region}
In contrast to the long-lived asymmetries at large distance from the star, the asymmetric structures from
debris from giant impacts that occur within a few au should be short-lived.
However, such giant impact debris is still potentially readily detectable \citep[e.g.,][]{Genda2015}.
This is because it takes only an asteroid's worth of dust to be detectable \citep[e.g.,][]{Kenyon2005},
and because this region is observed to be commonly empty of dust following dispersal of the protoplanetary disk
(see Fig.~\ref{fig:2}).

Moreover, the standard model for terrestrial planet formation involves many such impacts in the
timeframe 10-100\,Myr \citep{Kenyon2008}, with the collision that formed the Earth's moon being
a good example from the Solar System \citep[see][for a review of giant impacts]{Wyatt2016}.
The level of dust emission expected from debris created in the Moon-forming collision can
be calculated \citep{Jackson2012}, but there remains significant uncertainty in the duration of detectability.
For example, it is expected that $\sim 30$\% of the debris comes off as vapor that recondenses into cm-size grains
that are both bright and short-lived (a few 1000 orbits), while the remainder comes off as large boulders
that collisionally grind down and reaccrete onto the Earth and Venus over $\sim 15$\,Myr.
Thus the duration of detectability can be anywhere in the range $10^3-10^7$\,Myr depending on the details
of the boulder size distribution which is poorly constrained, as well as the dust optical properties.

Nevertheless, observationally it can be said that 3\% of nearby Sun-like stars in the age range 10-120\,Myr
show large levels of hot emission at 12\,$\mu$m \citep[which must originate from dust within a few au,][]{Kennedy2013}.
Warm emission traced at 24\,$\mu$m is also prevalent at this epoch \citep{Meng2017}.
In many cases it is hard to argue that this dust emission cannot be attributed to a massive young asteroid belt.
However, in others there is evidence to support a giant impact origin for the dust.
For example, in HD172555 the infrared spectrum shows spectral features indicative of the composition
of the dust, which is inferred to be primarily silica \citep[see Fig.~\ref{fig:5},][]{Lisse2009},
a composition expected to be produced in a hypervelocity impact.
For others, temporal evolution of the thermal emission is seen that may be attributed to the orbital
motion of a clump of vapour condensates \citep[see Fig.~\ref{fig:5},][]{Meng2014,Meng2015}.
These observations can potentially be used to characterise the aftermath of giant impacts, that
is, the size distribution of the fragments, their composition and their orbital properties.

%%%%%%%%%%%%%%%%%%%%%%%%%%%%%%%%%%%%%%%%%%%%%%%%%%%%%%%%%%%%%%
\subsection{Frequency of terrestrial planet formation}
Exoplanet surveys cannot yet tell us the fraction of stars that have Earth-like planets, or the fraction that
undergo terrestrial planet formation in the same way as the Solar System.
While the fraction of stars with Earth-like planets has been estimated by some authors to be just a few \%,
we know that super-Earth like planets (i.e., planets more massive than the Earth but residing much closer to the star)
are common, being found around 30-50\% of stars \citep[e.g., see][]{Winn2015}.
The frequency with which bright levels of hot dust are detected has the potential to say when (and where)
giant impacts occur, and so provides valuable information about the fraction of stars for which this mode
of planet formation operates.

Despite a few attempts \citep[e.g.,][]{Rhee2008,Wyatt2016}, it is hard to come to concrete conclusions yet from
the incidence of hot infrared excesses, primarily because such
conclusions rely on assumptions about the size distribution of the giant impact debris and the fraction of
infrared excesses that originate in giant impacts rather than asteroid belts (see above).
\citet{Wyatt2017} considered a different way of using the statistical information,
arguing that there is a sweet-spot in the mass and semimajor axis of the planet for which
its giant impact debris lasts longest above detectable levels (see orange contours on Fig.~\ref{fig:4} right).
This sweet-spot depends on details of the observation being performed, but generally requires planets
of a few $M_\oplus$ at a few au.
The inferred examples of giant impact debris found around A stars (like that in Fig.~\ref{fig:5} left)
are consistent with the expectation that this should be commonly found at temperatures appropriate for this sweet-spot,
whereas the giant impact debris found around Sun-like stars (like that in Fig.~\ref{fig:5} right) is inferred
to be at a fraction of an au.
This could suggest that terrestrial planet formation is rare beyond 1\,au around Sun-like stars, and that 
the examples of giant impact debris around such stars originate in the formation of super-Earths.
There are no examples of giant impact debris around M stars, which may be due to the short duration of detectability
for such emission, rather than an absence of giant impacts around M stars \citep{Wyatt2017}.

%%%%%%%%%%%%%%%%%%%%%%%%%%%%%%%%%%%%%%
\section{Conclusions}
\label{s:conc}
Debris disks provide valuable information about planet formation processes,
either by characterising the dispersal of the protoplanetary disk in which such processes take place,
by detailing the outcome of planetesimal and planet formation both in individual systems and on a population level,
or by bearing witness to such processes in action.

There is a growing number of detections of gas in debris disks.
The large gas masses and radial structure suggest that for some systems (particularly $10-40$\,Myr A stars)
primordial gas may persist after protoplanetary disk dispersal.
Lower levels of secondary gas are also now found to be common, requiring a volatile composition for the
planetesimals that is similar to Solar System comets.
Planetesimal formation at $>10$\,au occurs in at least 20\% of systems, with typically $>1M_\oplus$
left in $>10$\,km planetesimals at $>10$\,au.
It is hard to assess if the remaining 80\% of stars were inefficient at forming planetesimals, or
if their planetesimals have been depleted (perhaps because they only formed closer to the star).
Debris disks also provide clues to the level of stirring within the disks, which is sometimes low
($e<0.01$), and also suggest a dependence of planetesimal properties on stellar mass.

Many of the ways in which planets impose structure on debris disks have been characterised, and
these compare favourably with structures seen in the growing number of debris disk images.
In some cases the putative perturbing planet has been identified through other means confirming this
interpretation.
However, more such examples are needed before debris disk structure can be used with confidence to predict
unseen planets.
For example, regions that are empty of dust could be cleared by orbiting planets, but could equally
be regions of inefficient planetesimal formation.  
In the meantime, disk structures (particularly asymmetries) provide compelling evidence in favour of
planets and can be used to set constraints on any such planets' masses and orbits and in some cases their
dynamical histories. 
The first statistical evidence is also accumulating to link the properties of outer planetesimal belts
to the properties of inner planetary systems, though it is unclear as yet if this link is direct
(through planet-disk interactions) or indirect (through a common formation environment).

Ongoing planet formation processes are also expected to have a characteristic signature
in debris disk observations.
This could be through the slow growth of Pluto-sized bodies resulting in bright rings of debris at
10s of au up to Gyr, a long-lived asymmetric disk resulting from giant collisions in this outer disk,
or the release of large quantities of dust within a few au from the giant impacts expected between
planetary embryos during terrestrial planet formation.
The strongest evidence for ongoing planet formation comes from spectral and temporal observations
of a few young hot dust systems that characterise the aftermath of giant impacts.
Such detections have implications for the frequency of the formation of terrestrial planet and super-Earths
and its mode (i.e., whether this takes place through giant impacts).
However, uncertainties in the size distribution of giant impact debris and
the possibility that hot dust could also originate in massive asteroid belts
(rather than single impacts), means that no firm conclusions can yet be made.

%%%%%%%%%%%%%%%%%%%%%%%%%%%%%%%%%%%%%%
\section{Cross-References}
\begin{itemize}
\item{Andrews \& Birnstiel chapter on {\it Dust evolution in protoplanetary disks}}
\item{Bergin chapter on {\it Chemistry in protoplanetary disks}}
\item{Kalas chapter on {\it Fomalhaut's dust debris belt and eccentric planet}}
\item{Kral et al. chapter on {\it Circumstellar discs: What will be next?}}
\item{Morbidelli chapter on {\it Dynamical evolution of planetary systems}}
\item{Pudritz et al. chapter on {\it Connecting planetary composition to formation}}
\end{itemize}

%%%%%%%%%%%%%%%%%%%%%%%%%%%%%%%%%%%%%%
\bibliographystyle{spbasicHBexo}  %for bibtex
\bibliography{wyattbib} %for bibtex-example

\end{document}